\documentclass[11pt]{article}
\usepackage{graphicx}
\begin{document}

\title{Screening-Induced Transport at Finite Temperature in Bilayer Graphene}

\author{Min Lv and Shaolong Wan\thanks{Corresponding author.
Electronic address: slwan@ustc.edu.cn} \\
Institute for Theoretical Physics and Department of Modern Physics \\
University of Science and Technology of China, Hefei, 230026, {\bf
P. R. China}}

\date{\today}

\maketitle

\begin{abstract}
We calculate the temperature-dependent charge carrier transport of
bilayer graphene (BLG) impacted by Coulomb impurity scattering
within the random phase approximation. We find the polarizability is
equal to the density of states at zero momentum transfer and is
enhanced by a factor $\log{4}$ at large momentum transfer for
arbitrary temperature. The sharp cusp of static polarizability at
$q=2k_F$, due to the strong backward scattering, would be smooth by
the increasing temperatures. We also obtain the asymptotic behaviors
of conductivity of BLG at low and high temperature, and find it
turns from a two dimensional electron gas (2DEG) like linear
temperature metallic behavior to a single layer graphene (SLG) like
quadratic temperature insulating behavior as the temperature
increases.
\end{abstract}

PACS number(s): 81.05.Uw; 72.10.-d, 72.15.Lh, 72.20.Dp

\section{Introduction}

Since graphene, a two-dimensional single layer of graphite, is
fabricated \cite{Novoselov-1} which has attracted much attention
from both experimental and theoretical physicists. One important
experimental puzzle is that there is a so-called "minimum
conductivity" at the charge neutral (Dirac) point. Several early
theoretical work \cite{Katsnelson} calculated a universal $T=0$
minimum conductivity $\sigma_{min} = 4 e^2/\pi h$ at the Dirac point
in disorder-free graphene, but the experiments show that the
conductivity has a non-universal sample-dependent minimum
conductivity plateau ($\sim 4 e^2/h - 20 e^2/h$) around the Dirac
point. By using a random-phase approximation (RPA)-Boltzmann
formalism, this puzzle has been theoretically explained as the
result of carrier density fluctuations generated by charged
impurities in the substrate \cite{Hwang-1, Adam-1, Rossi}.
Therefore, the screened Coulomb scattering plays an important role
in understanding the transport properties of graphene, and several
corresponding theoretical researches \cite{Hwang-2, Ando, Wunsch,
Barlas} have been made.

While single layer graphene (SLG) has been widely studied from both
experimental and theoretical sides, bilayer graphene (BLG), which is
formed by stacking two SLG in Bernal stacking, as an other
significant carbon material, is attracting more and more attentions
\cite{McCann,Morozov,Novoselov-2} due to its unusual electronic
structure. BLG has a quadratic energy dispersion \cite{McCann}
similar to the regular two dimensional electron gas (2DEG) but its
effective Hamiltonian is chiral without bandgap similar to the SLG.
Although it is reported \cite{Zhang} that a widely tunable bandgap
has already successfully realized in BLG by using a dual-gate
bilayer graphene field-effect transistor and infrared
microspectroscopy, here we still ignore the bandgap in BLG
dispersion and keep the transport properties of such BLG with
tunable bandgap as a question studying in other paper.

Because of the role of screened Coulomb scattering by charged
impurities in understanding the transport properties of SLG, it is
significative to investigate the affection of screened Coulomb
scattering by charged impurities in BLG. The screening function of
BLG at zero temperature \cite{Hwang-3} and SLG at finite
temperature \cite{Hwang-4} have already been analytic
investigated, however, the BLG analytic form at finite temperature
and the corresponding temperature-dependent behaviors of
transport, which are the issues we will discuss in this article,
has not yet been provided. Although it is argued in Ref.
\cite{Adam-2} that any strong screening-induced temperature
dependence should not been anticipated in BLG resistivity and the
relatively strong collisional broadening effects would suppress
the small screening-induced temperature dependence due to the
small mobilities of current bilayer graphene samples, it is still
significative to investigate such screening-induced
temperature-dependent behavior for comparing the affections of
screened Coulomb scattering by charged impurities on transport
properties in BLG, SLG and 2DEG and representing how the BLG
behaves as the crossover from SLG to 2DEG.

This article is organized as the following. In Section 2, we present
the Boltzmann transport theory to calculate temperature-dependent
bilayer graphene conductivity. In Section 3, the
temperature-dependent screening function is investigated. In Section
4, we present the asymptotic behavior of conductivity at low and
high temperature, and numerical results obtained. The conclusion is
given in Section 5.

\section{Conductivity in Boltzmann Theory}

The BLG Hamiltonian has an excellent approximate form in the low
energy regime which can be written as \cite{McCann} (we set $\hbar
=1$ in this article)
\begin{eqnarray}
H_0 = - \frac{1}{2m} \left(
\begin{array}{cc}
0 & (k_x-ik_y)^2 \\
\\
(k_x + i k_y)^2 & 0
\end{array} \right), \label{2.1}
\end{eqnarray}
where $m = \gamma_1/(2 v_F^2) \approx 0.033 m_e$, $\gamma_1 \approx
0.39 eV$ is the interlayer coupling, and $v_F \approx 10^6 m/s$ is
the SLG Fermi velocity. The corresponding eigenstates of
Eq.(\ref{2.1}) are written as
\begin{eqnarray}
\Psi_{s{\vec{k}}} (\vec{r}) = \frac{1}{L}\exp(i \vec{k} \cdot
\vec{r}) F_{s{\vec{k}}}, \label{2.2}
\end{eqnarray}
with
\begin{eqnarray}
F_{s \vec{k}} =\frac{1}{\sqrt{2}} \left(
\begin{array}{c}
e^{- 2 i \theta_{\vec{k}}} \\
\\
s\end{array} \right), \label{2.3}
\end{eqnarray}
where $L^2$ is the area of the system, $s=+1$ and $-1$ denote the
conduction and valence bands, respectively, and
$\theta_{\vec{k}}=\arctan (k_y/k_x)$ is the polar angle of the
momentum $\vec{k}$. The corresponding energy is given by
$\epsilon_{s \vec{k}} = s k^2/{2m}$, and the BLG density of states
(DOS) is $N_0 = m g/{2 \pi}$ ($g = g_v g_s = 4$ is the total
degeneracy.) which is a constant for all energies and wave vectors.

When the electric field is small, the system is only slightly out of
equilibrium. To the lowest order in the applied electric field
$\vec{E}$, the distribution function can be written as $f_{s
\vec{k}} = f(\epsilon_{s \vec{k}}) + g_{s\vec{k}}$ where
$f(\epsilon_{s \vec{k}})$ is the equilibrium Fermi distribution
function and $g_{s \vec{k}}$ is the deviation proportional to
$\vec{E}$. The Boltzmann transport equation is given as
\begin{eqnarray}
\left(\frac{df_{s \vec{k}}}{dt} \right)_c = \frac{d \vec{k}}{dt}
\cdot \frac{\partial f(\epsilon_{s \vec{k}})}{\partial \vec{k}} = -
e \vec{E} \cdot \vec{v}_{s \vec{k}} \frac{\partial f}{\partial
\epsilon_{s \vec{k}}} = - \int \frac{d^2 k}{(2 \pi)^2} \left(g_{s
\vec{k}} - g_{s \vec{k'}} \right) W_{{s \vec{k}}, {s \vec{k'}}},
\label{2.4}
\end{eqnarray}
where $\vec{v}_{s \vec{k}} = s \vec{k}/m$ is the electron velocity,
\begin{eqnarray}
W_{{s \vec{k}}, {s \vec{k'}}} = 2 \pi n_i \mid \langle V_{{s
\vec{k}}, {s \vec{k'}}} \rangle \mid^2 \delta(\epsilon_{s \vec{k}} -
\epsilon_{s' \vec{k'}}), \label{2.5}
\end{eqnarray}
$n_i$ is the number of impurities per unit area, and $\langle V_{{s
\vec{k}}, {s \vec{k'}}} \rangle$ is the matrix element of scattering
potential with an average over configuration of scatterers. For
elastic impurity scattering, the interband processes ($s \neq s'$)
are forbidden. Under the relaxation-time approximation, we get
\begin{eqnarray}
g_{s \vec{k}} = - \tau(\epsilon_{s \vec{k}})e \vec{E} \cdot
\vec{v}_{s \vec{k}} \frac{\partial f(\epsilon_{s \vec{k}})}{\partial
\epsilon_{s \vec{k}}}, \label{2.6}
\end{eqnarray}
where the scattering time $\tau(\epsilon_{s \vec{k}})$ is given by
\begin{eqnarray}
\frac{1}{\tau(\epsilon_{s \vec{k}})} = 2 \pi n_i \int
\frac{d^2k'}{(2\pi)^2} \mid \langle V_{{s \vec{k}}, {s \vec{k'}}}
\rangle \mid^2 \left[ 1 - \cos{\theta_{\vec{k} \vec{k'}}} \right]
\delta(\epsilon_{s \vec{k}} - \epsilon_{s \vec{k'}}), \label{2.7}
\end{eqnarray}
and $\theta_{\vec{k} \vec{k'}}$ is the scattering angle between
$\vec{k}$ and $\vec{k'}$.

We know that the electrical current density is
\begin{eqnarray}
\vec{j} = g \int \frac{d^2k}{(2\pi)^2} e \vec{v}_{s \vec{k}}
f_{s\vec{k}}. \label{2.8}
\end{eqnarray}
Then we can get the electrical conductivity by using Eq.(\ref{2.6}),
\begin{eqnarray}
\sigma = \frac{N_0 e^2}{m} \int d\epsilon \tau(\epsilon) \epsilon
\left(- \frac{\partial f}{\partial \epsilon} \right). \label{2.9}
\end{eqnarray}
$f(\epsilon_k)$ is the Fermi distribution function $f(\epsilon_k) =
\{1 + \exp[ \beta (\epsilon_k - \mu) ] \}^{-1}$ where $\beta =
1/{k_BT}$ and  $\mu$ is the finite-temperature chemical potential.
At $T = 0$, $f(\epsilon) = \theta(\varepsilon_F - \epsilon)$(where
$\varepsilon_F \equiv \mu(T = 0)$), then we get the conductivity
formula $\sigma = \frac{e^2 v_F^2}{2} N_0 \tau(E_F)$ which has the
same form as the usual conductivity formula.

The matrix element of the scattering potential of randomly
distributed screened charge impurity in BLG is given as
\begin{eqnarray}
\mid \langle V_{{s \vec{k}}, {s \vec{k'}}} \rangle \mid^2 = \mid
\frac{v_i(q)}{\varepsilon(q)} \mid^2 \frac{1 + \cos{2\theta}}{2},
\label{2.10}
\end{eqnarray}
where $q = \mid \vec{k} - \vec{k'} \mid$, $\theta \equiv
\theta_{\vec{k} \vec{k'}}$,and $v_i (q) = 2 \pi e^2/(\kappa q)$ is
the Fourier transform of the potential of the charge impurity with a
background dielectric constant $\kappa$. The factor $(1 + \cos{2
\theta})/2$ is derived from the sublattice symmetry of BLG, while
this factor is replaced by $(1 + \cos{\theta})/2$ for SLG. The
finite-temperature RPA dielectric function can be written as
$\varepsilon(q) \equiv \varepsilon (q, T) = 1 + v_c(q)\Pi (q, T)$,
where $v_c(q)$ is the Coulomb potential and $\Pi(q, T)$ is the
irreducible finite-temperature polarization function. Then the
scattering time for energy $\epsilon_k$ of BLG is written as
\begin{eqnarray}
\frac{1}{\tau(\epsilon_k)} = \pi n_i \int \frac{d^2 k'}{(2 \pi)^2}
\mid \frac{v_i(q)}{\varepsilon(q)} \mid^2 \delta(\epsilon_k -
\epsilon_{k'})(1 - \cos{\theta})(1 + \cos{2 \theta}), \label{2.11}
\end{eqnarray}
comparing to the scattering time of SLG
\begin{eqnarray}
\frac{1}{{\tau(\epsilon_k)}_{SLG}} = \pi n_i \int \frac{d^2 k'}{(2
\pi)^2} \mid \frac{v_i(q)}{\varepsilon(q)} \mid^2 \delta(\epsilon_k
- \epsilon_{k'})(1 - \cos{\theta})(1 + \cos{\theta}), \label{2.12}
\end{eqnarray}
and the scattering time of 2DEG
\begin{eqnarray}
\frac{1}{{\tau(\epsilon_k)}_{2DEG}} = 2\pi n_i \int \frac{d^2 k'}{(2
\pi)^2} \mid \frac{v_i(q)}{\varepsilon(q)} \mid^2 \delta(\epsilon_k
- \epsilon_{k'})(1 - \cos{\theta}). \label{2.13}
\end{eqnarray}

We can find that formally the three formulas are almost the same
except the angular factor which arises from the sublattice symmetry,
$(1 + \cos{2\theta})/2$ for BLG, $(1 + \cos{\theta})/2$ for SLG and
$1$ for 2DEG. Actually the dielectric function $\epsilon(q)$ are
also different for the three systems, which would lead to different
scattering times in the three systems except for the angular
factors. They all have the same factor $(1 - \cos{\theta})$ which
weights the amount of scattering of the electron by the impurity and
always exists in Boltzmann transport formulism. This factor $(1 -
\cos{\theta})$ favors large-angle scattering events, which are most
important for the electrical resistivity of the regular 2D systems.
However, in SLG the large-angle scattering, in particular the $2k_F$
backward scattering, is suppressed by the factor $(1 +
\cos{\theta})$. In contrast to the regular 2D system, in SLG the
dominate contribution to the scattering time comes from the $k_F$
"right-angle" scattering ($i.e.$ $\theta=\pi/2$) but not the $2k_F$
backward scattering. Different from SLG, the $2k_F$ backward
scattering of the BLG is restored and even enhanced by the factor
$(1 + \cos{2 \theta})$ which arises from the sublattice symmetry of
BLG. Because of the restoral of the $2k_F$ backward scattering, many
theoretical approaches which fit the ordinary 2D systems can been
used for the BLG. Due to the qualitative similar, in some regimes
the temperature-dependent behavior of polarization function and the
transport properties in BLG are more similar to the 2DEG than the
SLG as we will show below.

\section{Temperature-Dependent Polarizability and Screening}

First let us consider temperature-dependent screening
\begin{eqnarray}
\epsilon(q, T) = 1 + \frac{2 \pi e^2}{\kappa q} \Pi(q, T),
\label{3.1}
\end{eqnarray}
where $\Pi(q, T)$ is the BLG irreducible finite-temperature
polarizability function, which is given by (calculated at $T=0$ in
Ref. \cite{Hwang-3} for BLG)
\begin{eqnarray}
\Pi(q, T) = - \frac{g}{L^2} \sum_{\vec{k} s s'} \frac{f_{s \vec{k}}
- f_{s' \vec{k'}}}{\varepsilon_{s \vec{k}} - \varepsilon_{s'
\vec{k'}}} F_{s s'}(\vec{k}, \vec{k'}), \label{3.2}
\end{eqnarray}
here $\vec{k'} = \vec{k} + \vec{q}$, $\varepsilon_{s \vec{k}} = s
k^2/{2m}$, and $F_{s s'}(\vec{k}, \vec{k'}) = (1 + s s' \cos{2
\theta})/2$ where $\theta \equiv \theta_{\vec{k} \vec{k'}}$, $f_{s
\vec{k}}$ is the Fermi distribution function $f_{s \vec{k}} = [\exp
\{\beta(\varepsilon_{s \vec{k}} -\mu) \} + 1]^{-1}$ where $\mu
\equiv \mu(T)$ is the finite-temperature chemical potential
determined by the conservation of the total electron density as
\begin{eqnarray}
\frac{T_F}{T} = F_0(\beta \mu) - F_0(- \beta \mu), \label{3.3}
\end{eqnarray}
where $T_F \equiv \varepsilon_F/{k_B}$ and
\begin{eqnarray}
F_n(x) = \int_{0}^{\infty} \frac{t^n dt}{\exp(t-x)+1}, \label{3.4}
\end{eqnarray}
It is easy to find that
\begin{eqnarray}
F_0 (x) = \log(1 + e^x), \label{3.5}
\end{eqnarray}
substitute it into Eq.(16), then we obtain
\begin{eqnarray}
\mu(T) = \varepsilon_F, \label{3.6}
\end{eqnarray}
which means that the chemical potential of BLG is
temperature-independent and very different from that of the SLG
and the regular 2D systems.

We rewrite the polarizability as
\begin{eqnarray}
\Pi(q, T) = \Pi_{intra}(q, T) + \Pi_{inter}(q, T), \label{3.7}
\end{eqnarray}
$\Pi_{intra}$ and $\Pi_{inter}$ indicate the polarization due to
intraband transition and interband transition, respectively, which
are given by
\begin{eqnarray}
\Pi_{intra}(q, T) = - \frac{g}{L^2} \sum_{\vec{k} s} \frac{f_{s
\vec{k}} - f_{s \vec{k'}}} {\varepsilon_{s \vec{k}} - \varepsilon_{s
\vec{k'}}} \frac{1 + \cos{2 \theta_{\vec{k} \vec{k'}}}}{2},
\label{3.8}
\end{eqnarray}
and
\begin{eqnarray}
\Pi_{inter}(q, T) = - \frac{g}{L^2} \sum_{\vec{k} s} \frac{f_{s
\vec{k}} - f_{- s \vec{k'}}} {\varepsilon_{s \vec{k}} -
\varepsilon_{- s \vec{k'}}} \frac{1 - \cos{2 \theta_{\vec{k}
\vec{k'}}}}{2}, \label{3.9}
\end{eqnarray}
where $\varepsilon_{s \vec{k}} = s k^2/{2m}$, $\vec{k'} = \vec{k} +
\vec{q}$, and
\begin{eqnarray}
\cos{2 \theta_{\vec{k} \vec{k'}}} = \frac{2 (k + q
\cos{\phi})^2}{\mid \vec{k} + \vec{q} \mid^2} - 1, \label{3.10}
\end{eqnarray}
here $\phi$ is an angle between $\vec{k}$ and $\vec{q}$. After
angular integration, we obtain
\begin{eqnarray}
\Pi_{intra}(q, T) = N_0 \int_{0}^{\infty} \frac{d k}{k^3}
\left[f(\varepsilon_k) + f(\varepsilon_k + 2\mu) \right] \left[k^2 -
\mid k^2 - q^2 \mid + \frac{(2 k^2 - q^2)^2}{q\sqrt{q^2 - 4 k^2}}
\theta(q - 2 k) \right], \label{3.11}
\end{eqnarray}
and
\begin{eqnarray}
\Pi_{inter}(q, T) &=& N_0 \int_{0}^{\infty} \frac{d k}{k^3} \left\{
\sqrt{4 k^4 + q^4} - k^2 - \mid k^2 - q^2 \mid \right. \nonumber \\
&&\left. - \left[ f(\varepsilon_k) + f(\varepsilon_k + 2\mu) \right]
\left[\sqrt{4 k^4 + q^4} - k^2 - \mid k^2 - q^2 \mid \right]
\right\}, \label{3.12}
\end{eqnarray}
here $N_0 = m g/2 \pi$ is the BLG density of states,
$f(\varepsilon)$ is the Fermi distribution function $f(\varepsilon)
= [\exp \{\beta(\varepsilon - \mu) \} + 1]^{-1}$. Then we have the
extrinsic BLG static polarizability at finite temperature as
\begin{eqnarray}
\Pi(q, T)&=& N_0 \int_{0}^{\infty} \frac{d k}{k^3} \left\{
\sqrt{4 k^4 + q^4} - k^2 - \mid k^2 - q^2 \mid \right. \nonumber \\
&&\left. + \left[ f(\varepsilon_k) + f(\varepsilon_k + 2 \mu)
\right] \left[2 k^2 - \sqrt{4 k^4 + q^4} + \frac{(2 k^2 - q^2)^2}{q
\sqrt{q^2 - 4 k^2}} \theta(q - 2 k) \right] \right\}. \label{3.13}
\end{eqnarray}
At high temperature $(T\gg T_F)$, Eq.(\ref{3.13}) can be written as
\begin{eqnarray}
\frac{\Pi(q, T)}{N_0} \approx 1 + \frac{q^2}{6 k_F^2} \frac{T_F}{T},
\label{3.14}
\end{eqnarray}
At low temperature $(T \ll T_F)$, Eq.(\ref{3.13}) can be written as
,
\begin{eqnarray}
\frac{\Pi (q, T)}{N_0} \approx g_0 (q) + \frac{\pi^2}{6}
\left(\frac{T}{T_F} \right)^2 g_1(q)~~~~~~~~~~~~(for~ q < 2 k_F),
\label{3.15}
\end{eqnarray}
and
\begin{eqnarray}
\frac{\Pi(q, T)}{N_0} \approx g_0(q) - f_0(q) + \frac{\pi^2}{6}
\left(\frac{T}{T_F} \right)^2 \left[g_1(q) - f_1(q)
\right]~~~~~~~~~~~~(for~ q > 2 k_F), \label{3.16}
\end{eqnarray}
with
\begin{eqnarray}
g_0(q) &=& \frac{1}{2 k_F^2} \sqrt{4 k_F^4 + q^4} - \log \left[ \frac{k_F^2 + \sqrt{k_F^4 + q^4/4}}{2 k_F^2} \right], \label{3.17} \\
f_0(q) &=& \frac{2 k_F^2 + q^2}{2 k_F^2 q} \sqrt{q^2 - 4 k_F^2} + \log \frac{q - \sqrt{q^2 - 4 k_F^2}}{q + \sqrt{q^2 - 4 k_F^2}}, \label{3.18}\\
g_1(q) &=& \frac{k_F^4 + q^4/2 - k_F^2 \sqrt{k_F^4 + q^4/4}}{k_F^2 \sqrt{k_F^4 + q^4/4}}, \label{3.19}\\
f_1(q) &=& \frac{(q^2 - _F^2)(q^4 - 5 q^2 k_F^2 + 2 k_F^4)}{k_F^2
q(q^2 -4 k_F^2)^{3/2}}. \label{3.20}
\end{eqnarray}
For $q= 2 k_F$, we have
\begin{eqnarray}
\frac{\Pi(q = 2 k_F, T)}{N_0}\approx C - \sqrt{\frac{\pi}{4}}(1 -
\sqrt{2}) \zeta(1/2) \left(\frac{T}{T_F} \right)^{1/2} -
\sqrt{\pi}(1 - \frac{\sqrt 2}{2}) \zeta(3/2) \left(\frac{T}{T_F}
\right)^{3/2}, \label{3.21}
\end{eqnarray}
where $C = \sqrt{5} - \log[(1 + \sqrt{5})/2]$, $\zeta(x)$ is
Riemann's zeta function. From above, we can give the doped BLG
polarizability at zero temperature which is same as the result
firstly obtained in Ref. \cite {Hwang-3},
\begin{eqnarray}
\frac{\Pi(q, T = 0)}{N_0} = g_0(q) - f_0(q) \theta(q -2 k_F).
\label{3.22}
\end{eqnarray}
The screened potential is
\begin{eqnarray}
U(q) = \frac{v(q)}{\epsilon(q)} = \frac{2 \pi e^2}{\kappa q [1 + v_c
\Pi(q)]} = \frac{2 \pi e^2}{\kappa (q + q_s)}, \label{3.23}
\end{eqnarray}
where $q_s(q, T) = q v_c(q) \Pi(q, T) = 2 \pi e^2\Pi(q, T)/\kappa =
q_{TF} \Pi(q, T)/N_0$ with $q_{TF} = m g e^2/\kappa$ being the
Thomas-Fermi screening wave vector of BLG. It is interesting to see
that in the $q\longrightarrow 0$ long wavelength limit, the $q_s(q,
T)$ of BLG is a constant value for all temperatures,
\begin{eqnarray}
q_s(q = 0, T) = q_{TF} = 4 r_s k_F, \label{3.24}
\end{eqnarray}
which is remarkably different from that of the SLG (see Eq.(29) and
(30) of Ref. \cite{Hwang-4}).
\begin{figure}
\includegraphics[width=8.5cm, height=6.0cm]{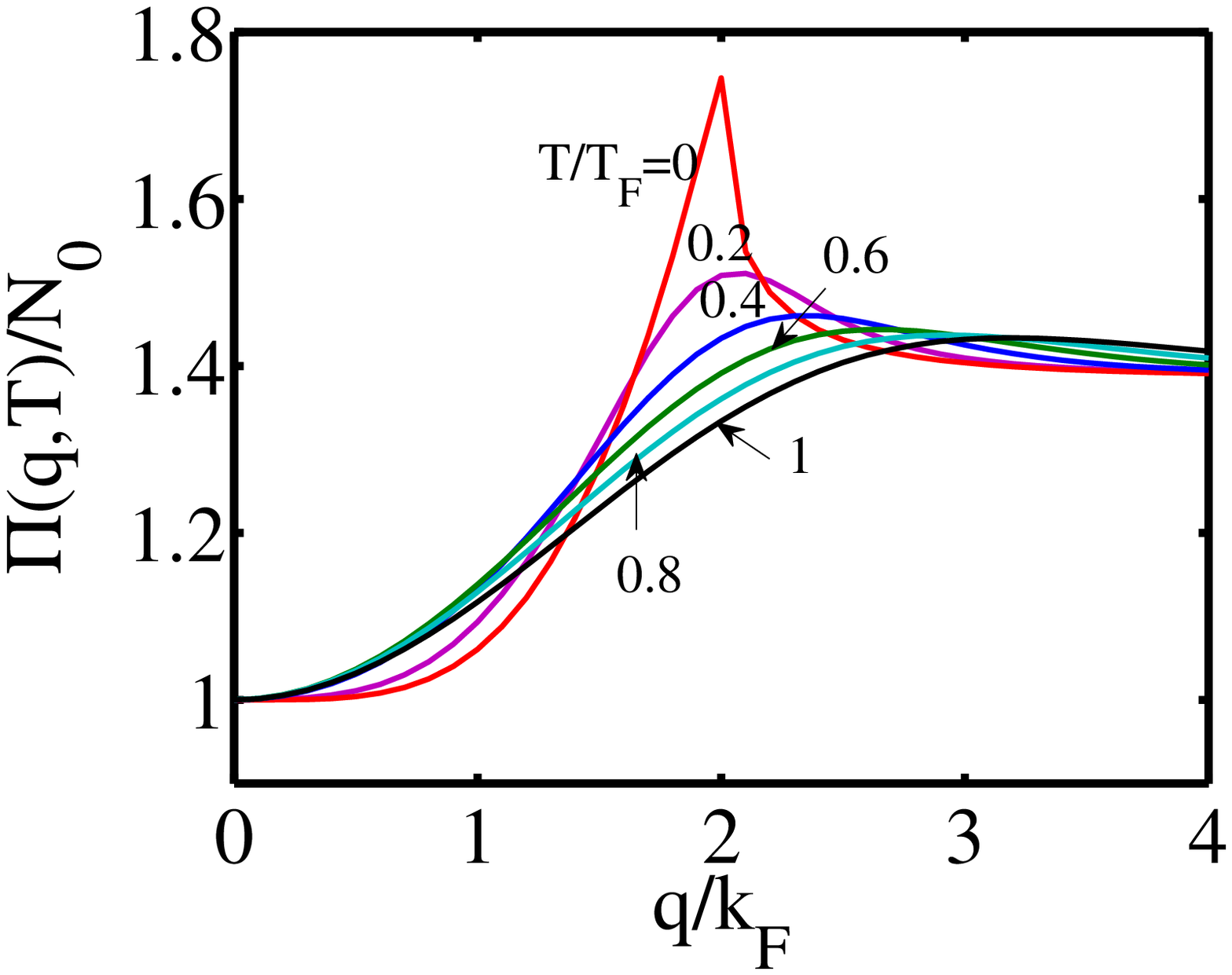}
\includegraphics[width=8.5cm, height=6.0cm]{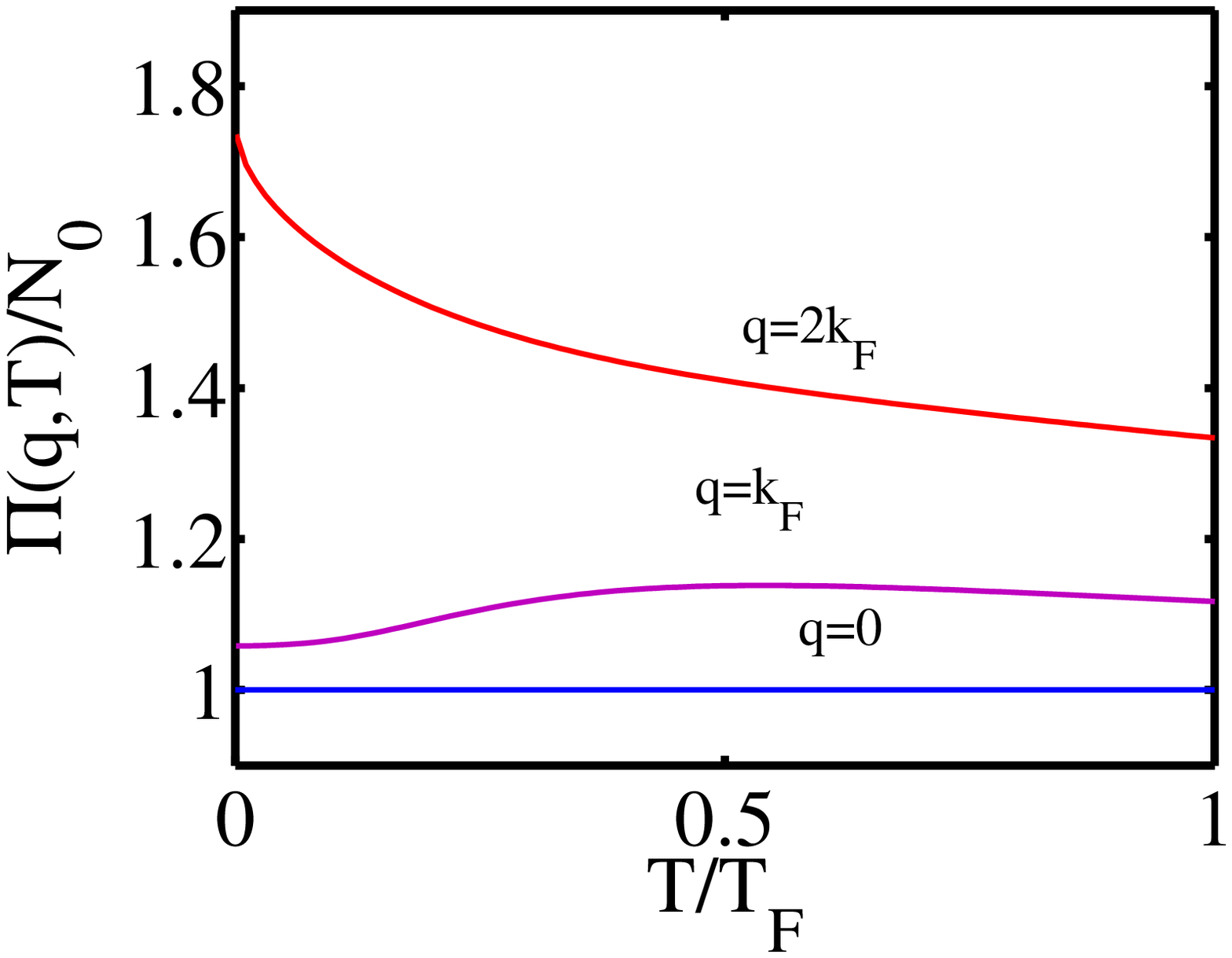}
\caption{Temperature-dependent BLG polarizability (a) as a function
of wave vector for different temperatures and (b) as a function of
temperature for different wave vectors. Here $N_0=mg/2\pi$.}
\label{fig.1}
\end{figure}

\begin{figure}
\includegraphics[width=8.5cm, height=6.0cm]{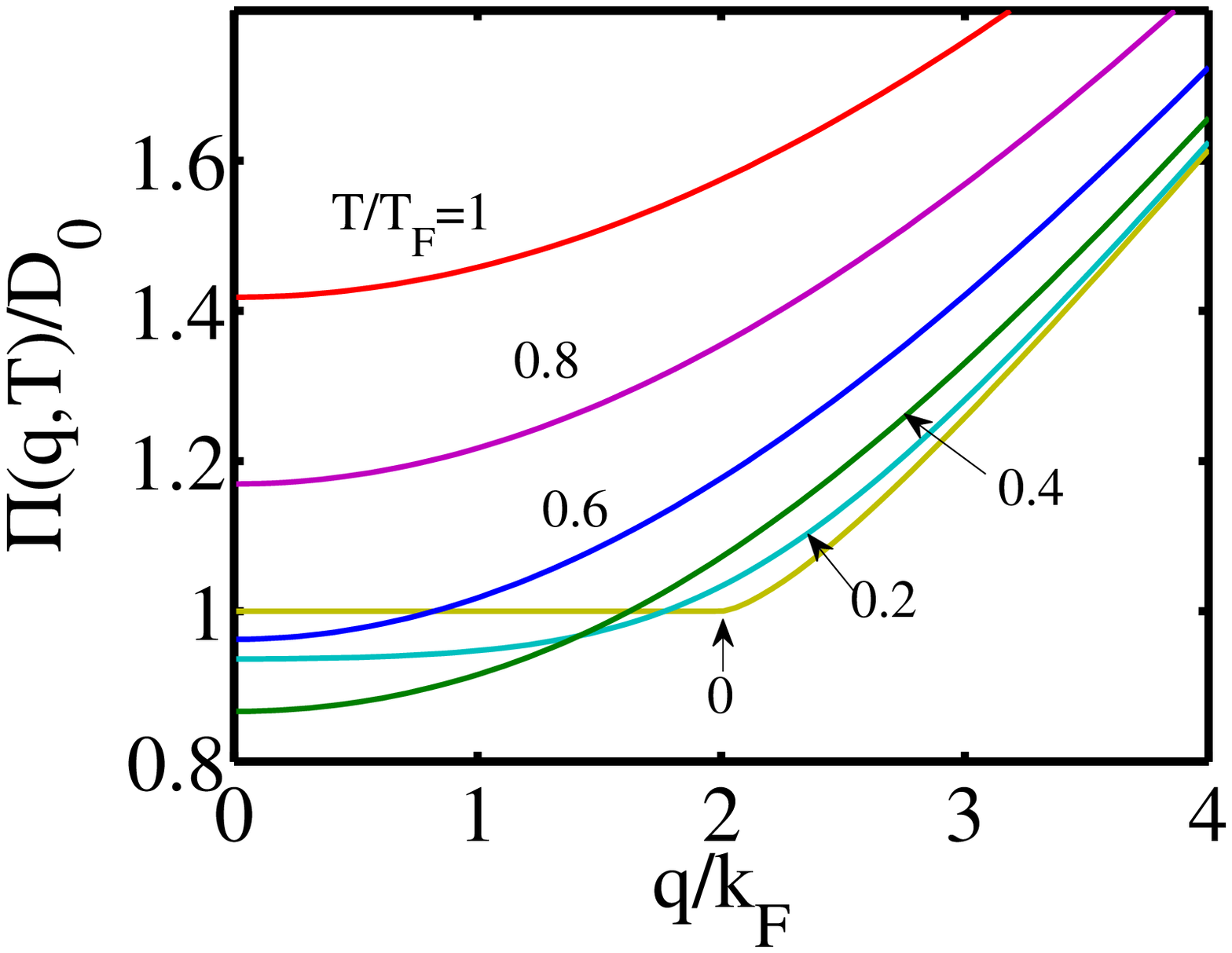}
\includegraphics[width=8.5cm, height=6.0cm]{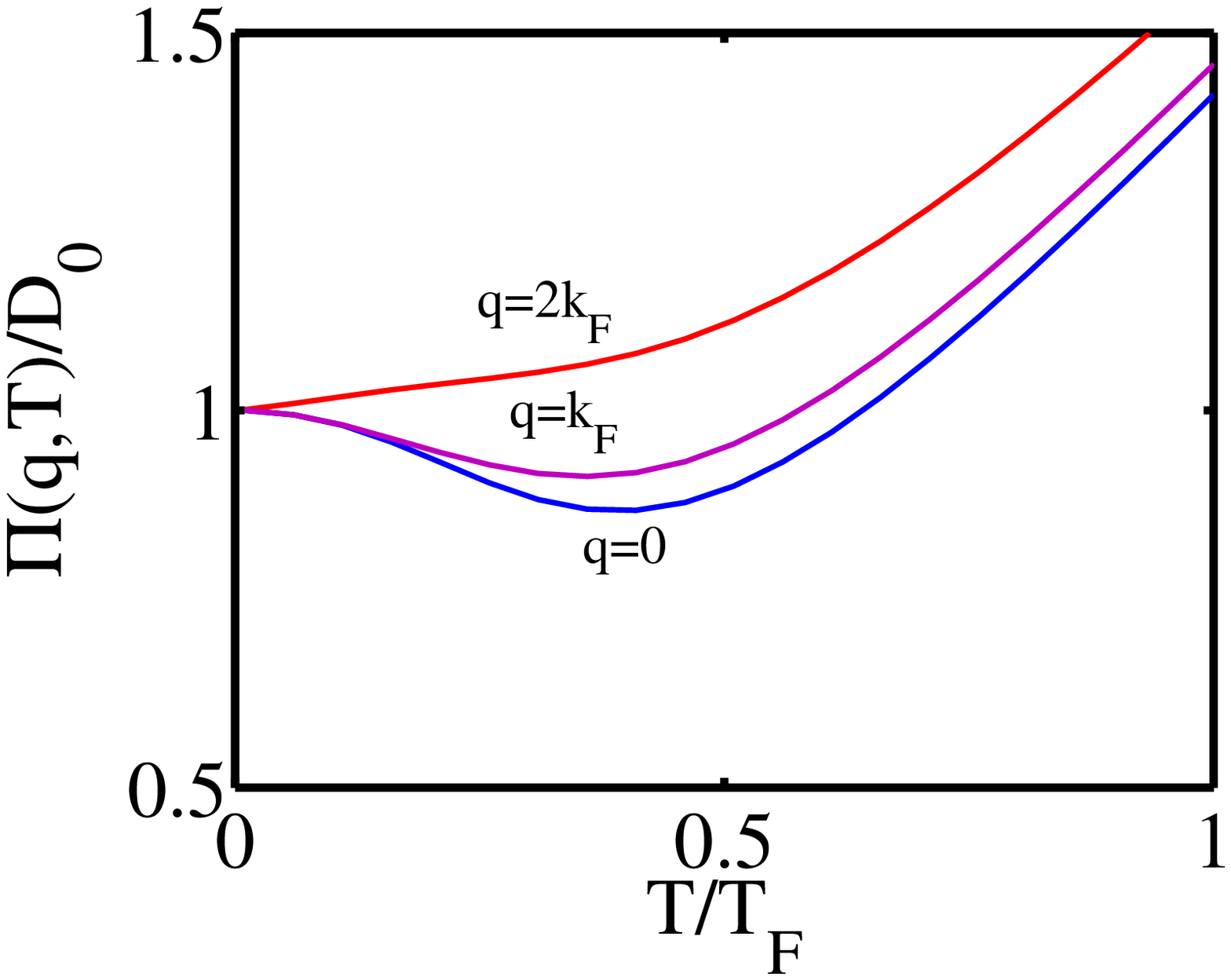}
\caption{Temperature-dependent SLG polarizability (a) as a function
of wave vector for different temperatures and (b) as a function of
temperature for different wave vectors. Here $D_0=gk_F/(2\pi v_F)$.}
\label{fig.2}
\end{figure}

\begin{figure}
\includegraphics[width=8.5cm, height=6.0cm]{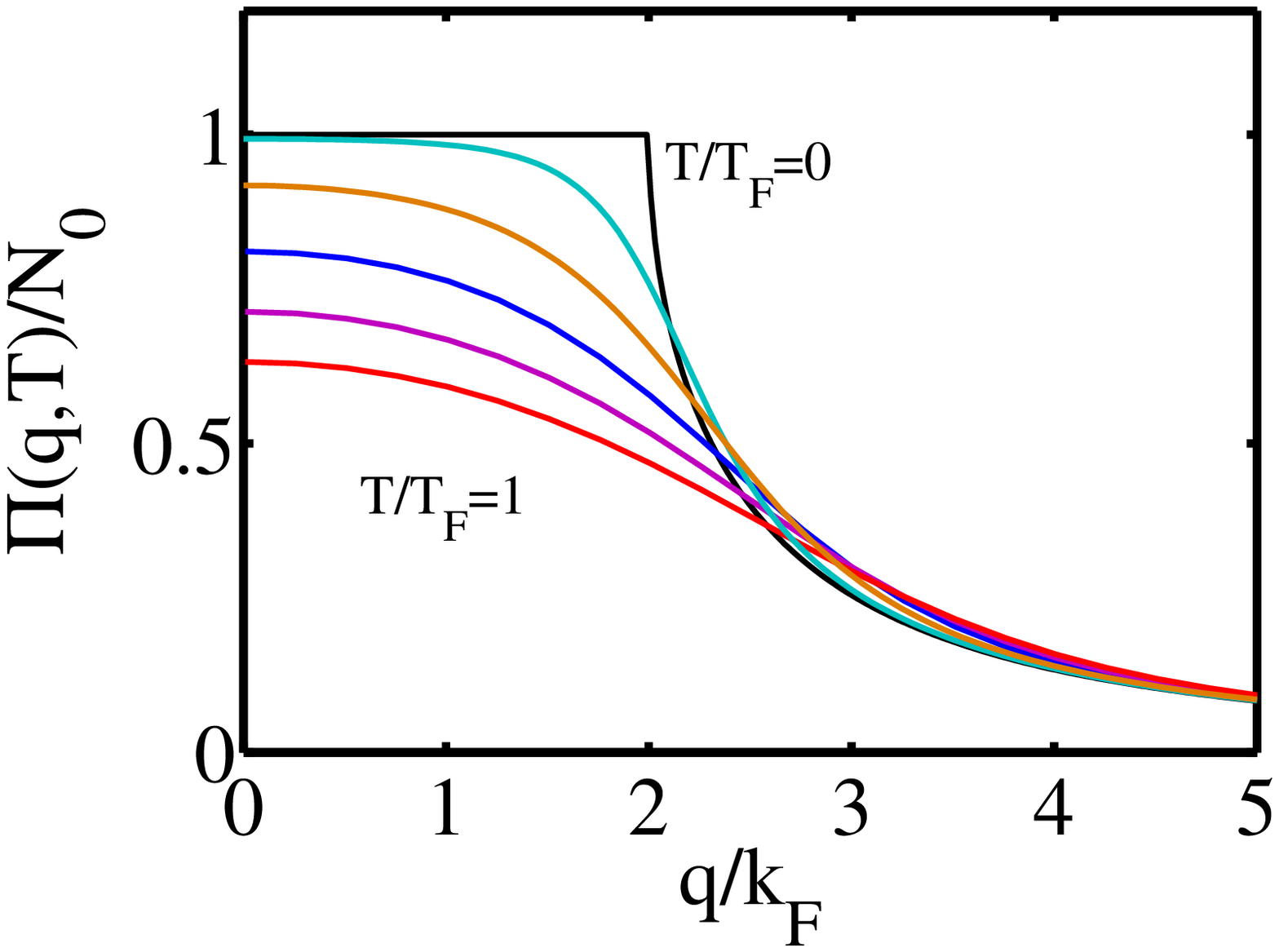}
\includegraphics[width=8.5cm, height=6.0cm]{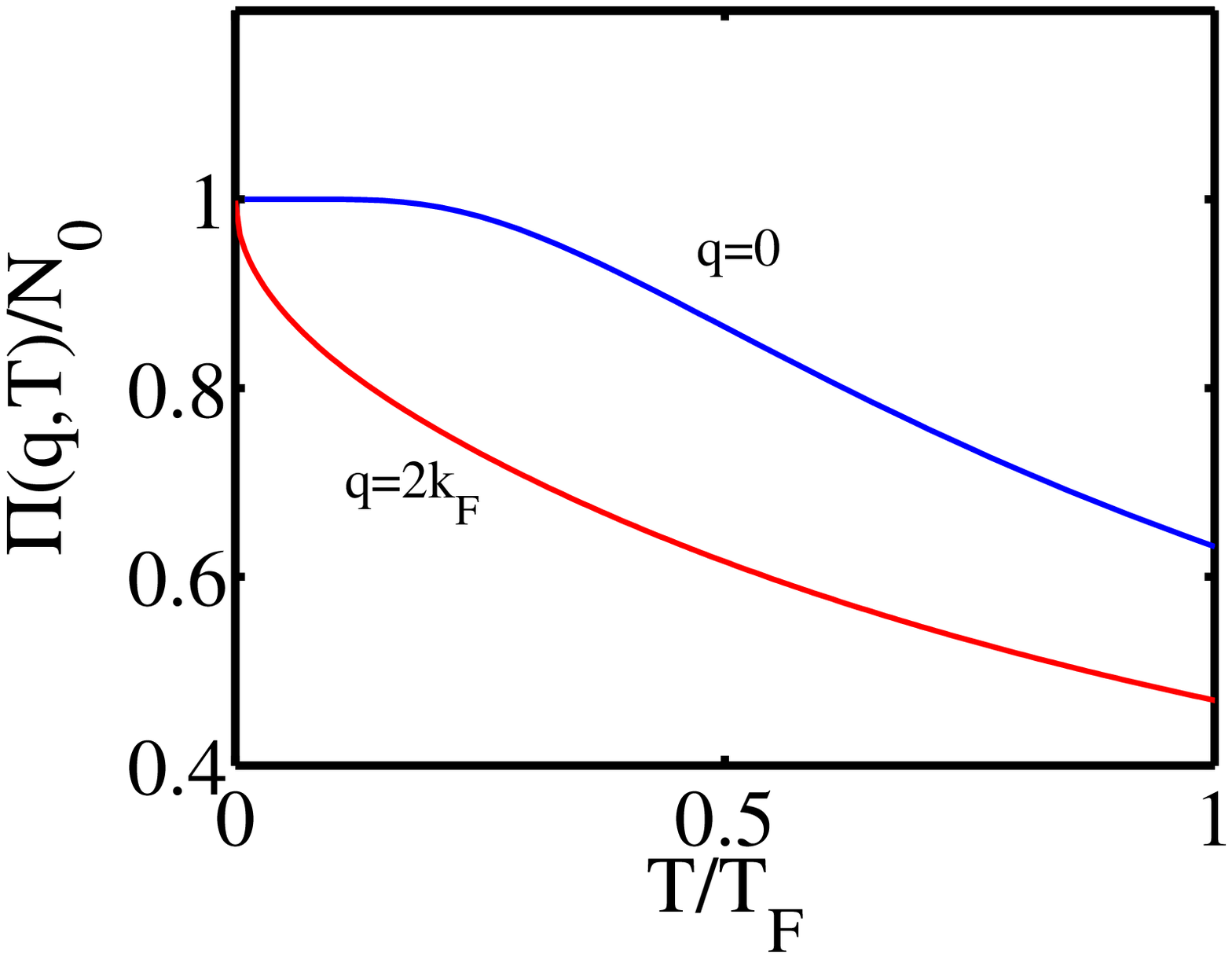}
\caption{The 2DEG polarizability function (a) as a function of wave
vector for several different temperatures
$T/T_F=0,0.2,0.4,0.6,0.8,1.0$ (top to bottom). (b) as a function of
temperature for different wave vectors. Here $N_0=mg/\pi$.}
\label{fig.3}
\end{figure}

The BLG finite-temperature polarizability $\Pi(q, T)$ as a function
of wave vector for different temperatures and as a function of
temperature for different wave vectors are shown in Fig.1 (a) and
(b), respectively. One novel phenomenon is that at $q = 0$, the BLG
polarizability equals to a constant value for all temperatures,
$i.e.$, $\Pi(q = 0, T) = N_0$. This is a qualitative difference
between BLG and SLG (or 2DEG) polarizability function, while the
latter $\Pi(q = 0, T)$ as a function of temperature changes notably
(as shown in Fig.2 for SLG and in Fig.3 for 2DEG). The reason is, at
$q = 0$, the intraband transition polarization $\Pi_{intra}(0, T) =
N_0$ while the interband transition polarization $\Pi_{inter}(0, T)
= 0$, $i.e.$, the interband transition is forbidden at zero momentum
transfer for all temperatures. The other remarkable phenomenon is
the BLG polarizabilty approaches a constant value $N_0\log{4}$ in
the large wave vector regime, arising from the fact that the
interband transition dominates over the intraband contribution in
the large wave vector limit. Therefore the dominative contribution
to the whole polarizability has a crossover from intraband
transition to interband transition for all temperatures. The weak
temperature-dependent behavior of polarizations in the small and
large wave vector regimes is a distinctive electronic property of
BLG. Different from that of BLG, as shown in Fig.2(a), the
polarizability of SLG increases monotonically for large $q$ stemming
from the domination of exciting electrons from the valence band to
the conduction band while the polarizability of 2DEG in large $q$
limit decreases as $1/{q^2}$ (as shown in Fig.3(a)).

In contrast to the SLG and the 2DEG, the polarizability of BLG as a
function of wave vector shows a nonmonotonicity, $i.e.$, the
polarizability at $T = 0$ is monotone increasing with $q$ in the
regime $[0, 2 k_F]$ and monotone decreasing in the regime larger
than $2 k_F$. In ordinary screened Coulomb scattering, the most
dominant scattering happens at $q = 2 k_F$, which gives rise to the
famous Friedel oscillations. Due to its sublattice symmetry, the $2
k_F$ backward scattering of SLG is suppressed and therefore there is
no singular behavior happen at $q = 2 k_F$. However, in BLG the $2
k_F$ backscattering is restored and even enhanced because of its
chirality, which leads to a sharp cusp and a discontinuous
derivation of polarizability at $T = 0$. We find that the
temperature dependence of BLG polarizability at $q = 2 k_F$ is
similar to that of 2DEG polarizability, both are much stronger than
that of SLG polarizability. Due to the strong temperature dependence
of the polarizability function at $q = 2 k_F$, as shown below, the
BLG would have a anomalously strong temperature-dependent
resistivity for $T \ll T_F$ which is similar to that of the regular
2D systems \cite{Sarma}. We also find the strong thermal suppression
of the singular behavior of polarizability at $q = 2 k_F$, which is
similar to that of the 2DEG.

Showing the difference between BLG and other 2D systems, we provide
the polarizability functions of SLG and regular 2DEG in the regimes
of $q=0$ and $q=2k_F$ in the low ($T \ll T_F$) and high ($T \gg
T_F$) temperature limits. For $T \ll T_F$,
\begin{eqnarray}
\Pi(q = 0, T) &\approx& \left\{
\begin{array}{ll} D_0[1 - \frac{\pi^2}{6}(\frac{T}{T_F})^2] &
( for \hspace{3mm}SLG ) \\
\\
N_0[1 - e^{- T_F/T}] & ( for \hspace{3mm} 2DEG )
\end{array} \right. \\
\label{3.25}
\Pi(q = 2 k_F, T) &\approx& \left\{
\begin{array}{ll} D_0 \left\{ \frac{\mu(T)}{E_F} + \sqrt{\frac{\pi
\mu}{2 E_F}} \left[1 - \frac{\sqrt{2}}{2} \right] \zeta
\left(\frac{3}{2} \right) \left(\frac{T}{T_F} \right)^{3/2}
\right\} & ( for \hspace{3mm} SLG ) \\
\\
N_0 \left[1 - \sqrt{\frac{\pi}{4}} \left(1 - \sqrt{2} \right) \zeta
\left( \frac{1}{2} \right) \left( \frac{T}{T_F} \right)^{1/2}
\right] & ( for \hspace{3mm} 2DEG ) \end{array} \right. \label{3.26}
\end{eqnarray}
here $D_0 = g E_F/2 \pi v_F^2$ and $N_0 = g m/2 \pi$ are the density
of states of SLG and regular 2DEG at Fermi level, respectively.
Comparing to the corresponding screening formula for BLG,
\begin{eqnarray}
\Pi(q = 0, T) &=& N_0, \\
\label{3.27}
\Pi(q = 2 k_F, T) &\approx&
N_0 [C - \sqrt{\frac{\pi}{4}}(1 - \sqrt{2})
\zeta(\frac{1}{2})(\frac{T}{T_F})^{1/2}]. \label{3.28}
\end{eqnarray}
We can find at $q = 2 k_F$ the zero-temperature value of
polarizability (normalized to the density of states at Fermi level)
of BLG is different from that of 2DEG, but their
temperature-dependent parts are both the same, which represents the
similarity of BLG and 2DEG.

For $T \gg T_F$,
\begin{eqnarray}
\Pi(q, T) &\approx& \left\{
\begin{array}{ll}
D_0 \frac{T}{T_F} [\ln{4} + \frac{q^2}{24 k_F^2}(\frac{T_F}{T})^2] &
( for \hspace{3mm} SLG ) \\
\\
N_0 \frac{T_F}{T} \left[1 - \frac{q^2}{6 k_F^2} \frac{T_F}{T}\right]
& ( for \hspace{3mm} 2DEG ) \end{array} \right. \label{3.29}
\end{eqnarray}
The corresponding high-temperature screening formula for BLG is
given by
\begin{eqnarray}
\Pi(q, T) \approx N_0 \left[1 + \frac{q^2}{6 k_F^2}\frac{T_F}{T}
\right]. \label{3.30}
\end{eqnarray}
In high temperature limit, the polarizability of BLG approaches a
constant value ($i.e.~ N_0$), which is very different from that of
SLG, where the static polarizability increases linearly with $T$,
and the regular 2DEG, where the polarizability falls as $1/T$. The
BLG shows an intermediate behavior between the SLG and the regular
2DEG.

\section{Conductivity Results}

\subsection{Analytic Asymptotic Results}

In this section, we study analytically the static conductivity of
the BLG in low and high temperatures limit. Firstly let us consider
the temperature dependence of conductivity in the low temperature
limit ($T \ll T_F$). Using Eq.(\ref{2.11}), the scattering time
$\tau(\varepsilon_F,T)$ at the Fermi level $\varepsilon_F$ in the
Born approximation is given as
\begin{eqnarray}
\frac{1}{\tau(\varepsilon_F, T)} = \frac{n_i}{2 \pi \varepsilon_F}
\int_{0}^{2k_F} d q \frac{q^2 [1 - 2 (q/2 k_F)^2]^2}{\sqrt{4 k_F^2 -
q^2}} \frac{v_i (q)^2}{\epsilon(q, T)^2}. \label{4.1}
\end{eqnarray}
In the low temperature limit, with Eq.(\ref{3.15}) and (\ref{3.16}),
we find the difference between the finite temperature polarizability
$\Pi(q,T)$ and the zero temperature polarizability $\Pi(q,T=0)$ is
just a second order $(\sim O(T^2))$ small quantity. Therefore, the
scattering time can be written as
\begin{eqnarray}
\frac{1}{\tau(\varepsilon_F, T)} \approx
\frac{1}{\tau(\varepsilon_F, T = 0)} + O(T^2), \label{4.2}
\end{eqnarray}
where
\begin{eqnarray}
\frac{1}{\tau(\varepsilon_F, T = 0)} = \frac{n_i}{2 \pi
\varepsilon_F} \int_{0}^{2 k_F} d q \frac{q^2 [1 - 2
(q/2k_F)^2]^2}{\sqrt{4 k_F^2 - q^2}} \frac{v_i(q)^2}{\epsilon(q, T =
0)^2}. \label{4.3}
\end{eqnarray}
It is easily seen that the $q \approx 2 k_F$ singularity dominate
the evaluation of the integral in Eq.(\ref{4.3}), thus we have
\begin{eqnarray}
\frac{1}{\tau(\varepsilon_F, T = 0)} &\approx& \frac{n_i}{2 \pi
\varepsilon_F} \left(\frac{2 \pi e^2}{\kappa} \right)^2 \frac{1}{[ 1
+ q_{TF} g_0(2k_F)/2 k_F]^2} \int_{0}^{1}d
x \frac{x^2 [1 - 2 x^2 ]^2}{\sqrt{1 - x^2}} \nonumber \\
&=& \frac{\pi}{g} \varepsilon_F \frac{n_i}{n} \frac{1}{[C + 2
k_F/q_{TF}]^2}, \label{4.4}
\end{eqnarray}
where $n$ is the electron density, $q_{TF} = m g e^2/\kappa$ is the
2D Thomas-Fermi screening wave vector, $C = g_0 (2 k_F) = \sqrt{5} -
\log[(1 + \sqrt{5})/2]$.

Considering the scattering time $\tau(\varepsilon,T=0)$ with energy
$\varepsilon=k^2/2 m $, we have
\begin{eqnarray}
\frac{1}{\tau(\varepsilon, T = 0)} = \frac{n_i}{2\pi \varepsilon}
\int_{0}^{2k} d q \frac{q^2 [ 1 - 2 (q/2k)^2]^2}{\sqrt{4 k^2 -
q^2}}\frac{v_i (q)^2}{\epsilon(q, T = 0)^2}, \label{4.5}
\end{eqnarray}
where
\begin{eqnarray}
\epsilon_1(q, T = 0) = 1 + v_c(q) N_0 g_0(q). \label{4.6}
\end{eqnarray}
Then we express $\epsilon(q, T = 0)$ as
\begin{eqnarray}
\epsilon(q, T = 0) = \epsilon_1 (q, T = 0) [1 - \frac{v_c(q) N_0
f_0(q)}{\epsilon_1(q, T = 0)} \theta(q -2 k_F)]. \label{4.7}
\end{eqnarray}
With Eq.(\ref{4.7}), we can express $1/\tau(\varepsilon,T=0)$ as
\begin{eqnarray}
\frac{1}{\tau(\varepsilon, T = 0)} \approx
\frac{1}{\tau_0(\varepsilon, T = 0)} + \frac{1}{\tau_1(\varepsilon,
T = 0)}, \label{4.8}
\end{eqnarray}
where
\begin{eqnarray}
\frac{1}{\tau_0(\varepsilon, T = 0)} = \frac{n_i}{2 \pi \varepsilon}
\int_{0}^{2k} d q \frac{q^2 [1 - 2(q/2 k)^2]^2}{\sqrt{4 k^2 - q^2}}
\frac{v_i(q)^2}{\epsilon_1 (q, T = 0)^2}, \label{4.9}
\end{eqnarray}
and
\begin{eqnarray}
\frac{1}{\tau_1(\varepsilon, T = 0)} = \frac{n_i}{\pi \varepsilon}
\int_{2 k_F}^{2 k} d q \frac{q^2 [1 - 2(q/2k)^2]^2}{\sqrt{4 k^2 -
q^2}}\frac{v_i(q)^2}{\epsilon_1(q, T = 0)^2} \frac{v_c(q) N_0
f_0(q)}{\epsilon_1 (q, T = 0)} \theta(\varepsilon - \varepsilon_F).
\label{4.10}
\end{eqnarray}
For $\mid (\varepsilon - \varepsilon_F)/\varepsilon_F \mid \ll 1$,
we can write
\begin{eqnarray}
\frac{1}{\tau_0(\varepsilon, T = 0)} \approx \frac{1}
{\tau(\varepsilon_F, T = 0)[1 + A(\varepsilon -
\varepsilon_F)/\varepsilon_F]}, \label{4.11}
\end{eqnarray}
with $A = -\varepsilon_F\tau(\varepsilon_F, T = 0) \partial
[1/\tau(\varepsilon_F, T = 0)]/\partial \varepsilon_F$. From
Eq.(\ref{4.10}) we use the same trick as that of Eq.(\ref{4.4}) and
obtain
\begin{eqnarray}
\frac{1}{\tau_1(\varepsilon, T = 0)} = \frac{n_i}{\pi
\varepsilon}(\frac{v_i(2 k_F)}{\epsilon_1(2 k_F, T = 0)})^2
\frac{v_c(2 k_F)N_0}{\epsilon_1(2 k_F, T = 0)} I_1
\theta(\varepsilon - \varepsilon_F), \label{4.12}
\end{eqnarray}
where
\begin{eqnarray}
I_1 = \int_{2 k_F}^{2 k} d q \frac{q^2 [1 - 2 (q/2k)^2]^2} {\sqrt{4
k^2 - q^2}}[\frac{2 k_F^2 + q^2}{2 k_F^2 q} \sqrt{q^2 - 4 k_F^2} +
\log{\frac{q - \sqrt{q^2 - 4 k_F^2}}{q + \sqrt{q^2 - 4 k_F^2}}}],
\label{4.13}
\end{eqnarray}
and for $\mid (\varepsilon - \varepsilon_F)/\varepsilon_F \mid \ll
1$, we have
\begin{eqnarray}
I_1 = \pi k_F^2 \frac{\varepsilon - \varepsilon_F}{\varepsilon_F} +
O((\frac{\varepsilon - \varepsilon_F}{\varepsilon_F})^2).
\label{4.14}
\end{eqnarray}
With Eq.(\ref{4.8})-(\ref{4.14}), the energy dependent conductivity
at zero temperature is given as
\begin{eqnarray}
\sigma(\varepsilon, T = 0) &=&
\frac{e^2 v_F^2}{2} N_0 \tau(\varepsilon, T = 0) \nonumber \\
 &\approx& \frac{e^2 v_F^2}{2} N_0 \tau(\varepsilon_F, T = 0) \left[1 + A \frac{\varepsilon - \varepsilon_F}
{\varepsilon_F} - \frac{\tau_0 (\varepsilon, T =
0)}{\tau_1(\varepsilon, T = 0)} \right] \label{4.15}
\end{eqnarray}
where
\begin{eqnarray}
\frac{\tau_0(\varepsilon, T = 0)}{\tau_1 (\varepsilon, T = 0)} =
\frac{4 q_{TF}} {2 k_F + C q_{TF}} \frac{\varepsilon -
\varepsilon_F}{\varepsilon_F} \theta(\varepsilon - \varepsilon_F).
\label{4.16}
\end{eqnarray}
Using the Kubo-Greenwood formula \cite{Ziman}
\begin{eqnarray}
\sigma(\varepsilon_F, T) = \frac{1}{4 k_BT} \int_0^{\infty}d
\varepsilon \frac{\sigma(\varepsilon, T = 0)}{\cosh^2 [(\varepsilon
- \varepsilon_F)/2 k_BT]}, \label{4.17}
\end{eqnarray}
and substituting Eq.(\ref{4.15}) and (\ref{4.16}) into
Eq.(\ref{4.17}) with a consideration of Eq.(\ref{4.4}), we obtain
the analytic asymptotic behavior of BLG conductivity at low
temperature as following
\begin{eqnarray}
\sigma(T\ll T_F) = \sigma_0^{2 D} \left( 1 - \frac{4 \log{2}}{C +
1/q_0} \frac{T}{T_F} \right), \label{4.18}
\end{eqnarray}
where $q_0 = q_{TF}/2k_F$ and $\sigma_0^{2D} = e^2 v_F^2 N_0
\tau(\varepsilon_F, T = 0)/2$.

For high temperature limit, substituting Eq.(\ref{3.3}) into Eq.
(\ref{2.9}), we have
\begin{eqnarray}
\sigma(T\gg T_F)=\sigma_1^{2D}\frac{\pi^2}{6} \left(\frac{T}{T_F}
\right)^2 \left(1+C_h q_0\sqrt{\frac{T_F}{T}}\right), \label{4.19}
\end{eqnarray}
with $C_h = 3.57$ and $\sigma_1^{2D} = (e^2/h)(n/n_i)(g^2/\pi
q_0^2)$.

We show our numerical results of the conductivity and analytic
asymptotic result of Eq.(\ref{4.18}) in Fig.4, and find that they
are excellently agreement with the numerical results in the low
temperature limit.

It is significative to compare the conductivity temperature
behaviors of SLG, BLG and 2DEG. For SLG, the asymptotic low and
high temperature behaviors of conductivity are given by
\cite{Hwang-4}
\begin{eqnarray}
\sigma(T \ll T_F) = \sigma_0 \left[1-C_0\frac{\pi^2}{3}
\left(\frac{T}{T_F} \right)^2 \right], \label{4.20}
\end{eqnarray}
\begin{eqnarray}
\sigma(T \gg T_F) = \sigma_0 \frac{16I_0}{\pi} \left[ 4\log({2})r_s
\right]^2 \left(\frac{T}{T_F} \right)^2, \label{4.21}
\end{eqnarray}
here $\sigma_0 \equiv \sigma (T = 0)$, $C_0 \sim o(1)$ and
$I_0=0.034$. For 2DEG as found in Si MOSFETs and GaAs
heterostructures, the asymptotic low and high temperature behaviors
of conductivity are written as \cite{Sarma-2}
\begin{eqnarray}
\sigma(T \ll T_F) \approx \sigma_0^{2D} \left[ 1 - C_1
\left(\frac{T}{T_F} \right) \right], \label{4.22}
\end{eqnarray}
\begin{eqnarray}
\sigma(T\gg T_F) \approx \sigma_1^{2D} \left[\frac{T}{T_F} + \frac{3
\sqrt{\pi} q_0}{4} \sqrt{\frac{T_F}{T}} \right].
\end{eqnarray}
here $\sigma_0^{2D} \equiv \sigma (T = 0)$, and $C_1 = 2 q_0/(1 +
q_0)$, $\sigma_1^{2D} = (e^2/h)(n/n_i)(g^2/\pi q_0^2)$ where $q_0 =
q_{TF}/2k_F$.

Now let us compare the BLG temperature dependence with the SLG and
the regular parabolic 2D systems. First, for $T \ll T_F$, all the
three systems show metallic temperature-dependent behaviors, but
their strengthes of temperature dependence are different. BLG and
the parabolic 2D system both have strong linear temperature
dependence while SLG has a weak quadratic temperature dependence.
Second, for high temperature limit $T \gg T_F$, BLG represents a
quadratic temperature-dependent behavior which is similar to SLG,
compared with the linear temperature dependence in the parabolic 2D
system. Therefore, in the low temperature regime, the
temperature-dependent transport of BLG is qualitatively similar to
that of the parabolic 2D system, but as the temperature increasing,
BLG is getting more and more similarity with SLG. The transport
property of BLG as the intermediate between SLG and the regular 2DEG
has been shown here.

\begin{figure}
\includegraphics[width=8.5cm, height=6.0cm]{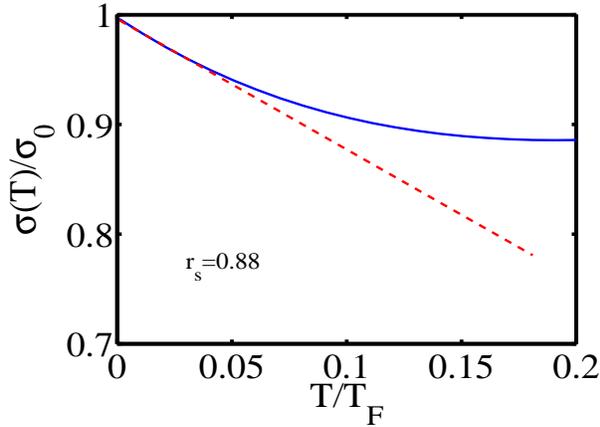}
\caption{Numerical results of temperature-dependent conductivity of
BLG in the low temperature region and its asymptotic form given by
Eq.(\ref{4.18}). The asymptotic form fits well in the regime $T\sim
T_F[0,0.04]$. Here $\sigma_0=e^2v_F^2N_0\tau(\varepsilon_F,T=0)/2$.
} \label{fig.4}
\end{figure}

\subsection{Numerical Results}

We show the numerical results of resistivities obtained from
Eq.(\ref{2.9}) as a function of temperature for different $r_s$
values in Fig.5. $r_s = 0.88$ corresponds to substrate-mounted
($\kappa = 4$) bilayer graphene and $r_s = 3.51$ corresponds to
suspended ($\kappa = 1$) bilayer graphene. It is found that the
numerical results of BLG resistivity show metallic behavior at low
temperature and insulating behavior at high temperature, which is
the same as SLG and the regular 2D systems. Unlike graphene the
scaled temperature-dependent resistivity of BLG has a relatively
strong dependence on $r_s$ which is similar to that of the regular
2D systems. Because of the strong $q = 2 k_F$ backward scattering
occurring in these two systems, this similarity also represents in
the low temperature regime, where both of them have strong linear
$T$ metallic behaviors with slopes $4 \log{2}/(C + 1/q_0)$ for BLG
and $2/(1 + 1/q_0)$ for 2DEG. With temperature increasing, the BLG
temperature-dependent behavior of resistivity is changing from
2DEG-like to SLG-like (falling off rapidly as $\sim 1/T^2$).

For comparison we show the calculated temperature-dependent
resistivity of SLG and ordinary 2D systems for different interaction
parameters $r_s$ in Fig.6 (for SLG) and Fig.7 (for ordinary 2D
system). These two figures come from Ref.\cite {Hwang-4}. It is
intuitively seen that at low temperature the linear $T$ behavior of
BLG is similar to that of ordinary 2D system. However, the linear
$T$ regime is rather weak and narrow for BLG, which is about $T \sim
T_F [0, 0.04]$ for $r_s = 0.88$, while the linear $T$ regime for
ordinary 2D system is relatively strong and broad, which extends
from zero temperature to about $0.5 T_F$ for $r_s = 2.6$. Therefore
in BLG, this screening-induced linear $T$ behavior is easily
suppressed by other effects.

Since the Wigner-Seitz radius $r_s$, representing the strength of
electron-electron interaction, is reasonably small ($r_s \approx
0.88$ for carrier density $n \sim 10^{12} cm^{-2}$), as shown in
Fig.5 the temperature dependence arising from screening is rather
weak in the BLG(the resistivity of BLG for $r_s = 3.51$ decreases
just about $14$ percents from $T/T_F = 0$ to $T/T_F = 1$ while the
increase of resistivity of 2DEG for $r_s = 3.7$ exceeds $100$
percents from $T/T_F = 0$ to $T/T_F = 1$). The dimensionless
temperature $T/T_F$ is also rather small because of the relatively
high Fermi temperature in BLG ($T_F \sim 400K$ for $n \sim 10^{12}
cm^{-2}$). Therefore, as investigated experimentally in Ref.
\cite{Morozov} and argued in Ref. \cite {Adam-2}, the strong
collisional broadening effects due to the very small mobilities of
current BLG samples would suppress the weak screening-induced
temperature dependence that we calculated at low temperature ($i.e.,
T \ll T_F$) and there would be no much temperature dependence in the
low temperature resistivity. We hope our results to be tested in
future experiments.

We show the temperature-dependent conductivity of BLG for different
temperatures calculated as a function of carrier density in Fig.8.
It shows that the conductivity increases in the low density regime
and decrease in the high density regime as the temperature
increases, representing a non-monotonic behavior of the
conductivity.

In this article, we have assumed an ideal 2D BLG electron gas and
ignored the distance $d$ between bilayer graphene and the charged
impurity located at the substrate, which would make the form of
potential of the charged impurity become $2\pi e^2 e^{-q d}/\kappa(q
+ q_{TF})$. We also have assumed a homogeneous carrier density
model, and therefore our theory is quantitatively correct only in
the relatively high-density regime where the spatially inhomogeneous
effects arising from charged impurity-induced electron-hole puddles
are weak. For just comparing theoretically the screening-induced
temperature-dependent behaviors of different 2D systems ($i.e.$,
BLG, SLG and 2DEG), we do not take the level broadening effects due
to impurity-scattering into account.

\begin{figure}
\includegraphics[width=8.5cm, height=6.0cm]{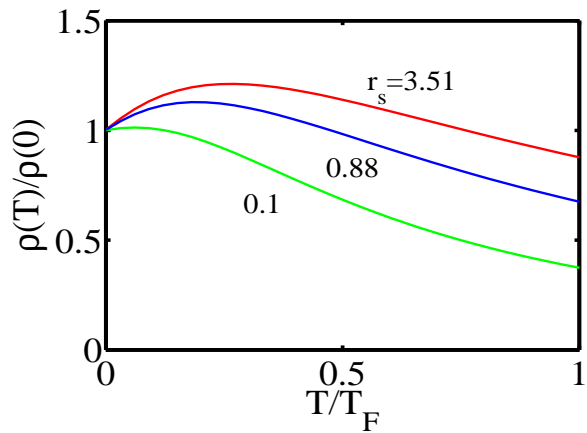}
\caption{Numerical results of resistivities obtained from
Eq.(\ref{2.9}) as a function of temperature $T/T_F$ for different
$r_s = 3.51, 0.88, 0.1$ (from top to bottom). $r_s=0.88 (3.51)$
corresponds to substrate-mounted (suspended) BLG. As $r_s$ increases
the metallic behavior becomes stronger.} \label{fig.5}
\end{figure}

\begin{figure}
\includegraphics[width=8.5cm, height=6.0cm]{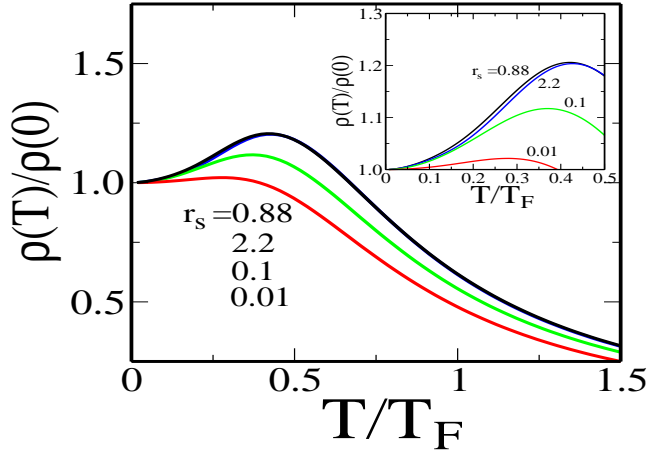}
\caption{Numerical results of resistivities as a function of
temperature $T/T_F$ for different $r_s=0.88$, 2.2, 0.1, 0.01 (from
top to bottom). $r_s=0.88$ (2.2) corresponds to graphene on the
SiO$_2$ substrate (in vacuum). Inset shows the magnified view in the
low temperature limit $T < 0.5T_F$. This figure comes from Ref.
\cite{Hwang-4}.} \label{fig.6}
\end{figure}

\begin{figure}
\includegraphics[width=8.5cm, height=6.0cm]{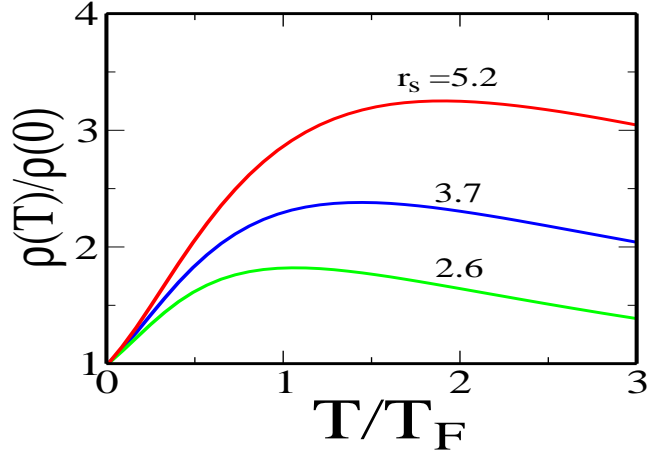}
\caption{ $\rho(T)/\rho(0)$ of an ordinary 2D system for different
$r_s$ values as a function of temperature. As $r_s$ increases the
metallic behavior becomes stronger. Compering to Fig. 5, the
metallic behavior of the ordinary 2D system is much stronger than
that of the BLG. This figure comes from Ref. \cite{Hwang-4}.}
\label{fig.7}
\end{figure}

\begin{figure}
\includegraphics[width=8.5cm, height=6.0cm]{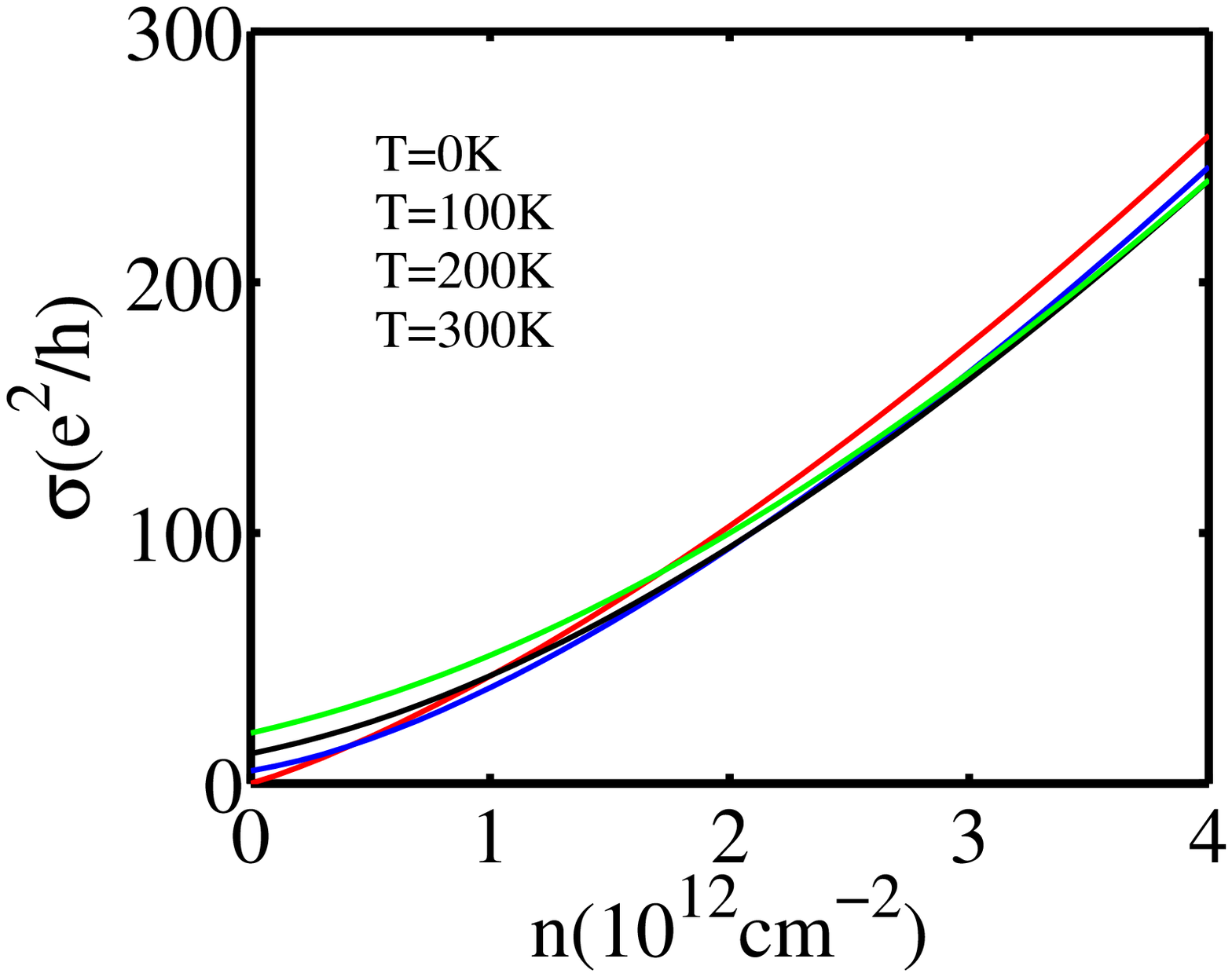}
\caption{ Calculated conductivity for different temperatures $T=0,
100, 200, 300 K$ (bottom to top in low density regime) as a function
of density. Here we use $r_s=0.88$ and an impurity density
$n_i=5\times 10^{11}cm^{-2}$.} \label{fig.8}
\end{figure}

\section{Conclusions}

In this article, we calculate the static wave vector polarizability
of doped bilayer graphene at finite temperature under the RPA. We
find that for all temperatures, the BLG static screening is equal to
its density of states $N_0$ at zero momentum transfer and enhanced
by a factor of $\log{4}$ at large momentum transfer. Due to the
enhanced $q = 2 k_F$ backward scattering arising from the chirality
of the BLG, a strong cusp of polarizability occurs at $q = 2 k_F$
for zero temperature but strongly thermal suppressed as temperature
increases. Using a microscopic transport theory for BLG conductivity
at finite temperature, we also obtain the asymptotic low and high
temperature behaviors of conductivity for BLG, and find it has a
linear temperature metallic behavior similar to the regular 2D
system at low temperature and a quadratic temperature insulating
behavior similar to the SLG. This crossover from 2DEG-like behavior
to SLG-like behavior as temperature increases represents the unique
transport properties of BLG as intermediate between the SLG and the
regular 2DEG.

\section*{Acknowledgement}

M. Lv acknowledges Mengsu Chen for his help. This work is
supported by NSFC Grant No.10675108.

\end{document}